\documentclass[aps,twocolumn,showpacs,superscriptaddress]{revtex4}
\usepackage{graphicx,color}
\usepackage{amsmath}
\usepackage{amssymb}

\usepackage{amsfonts}
\usepackage{bm}

\newcommand{\ot}{{\,\otimes\,}}
\newcommand{{\Cd}}{{\mathbb{C}^d}}

\def\oper{{\mathchoice{\rm 1\mskip-4mu l}{\rm 1\mskip-4mu l}%
{\rm 1\mskip-4.5mu l}{\rm 1\mskip-5mu l}}}
\def\<{\langle}
\def\>{\rangle}

\newtheorem{example}{Example}

\begin{document}
\title{General form of quantum evolution}
\author{Dariusz Chru\'sci\'nski and Andrzej Kossakowski}
\affiliation{Institute of Physics, Nicolaus Copernicus University \\
Grudzi\c{a}dzka 5/7, 87--100 Toru\'n, Poland}

\begin{abstract}

We propose a complete treatment of a local in time dynamics of open
quantum systems. In this approach Markovian evolution turns out to
be a special case of a general non-Markovian one. We provide a
general representation of the local  generator which generalizes
well known Lindblad representation for the Markovian dynamics. It
shows that the structure of non-Markovian generators is highly
intricate and the problem of their classification is still open.
Simple examples illustrate our approach.

\end{abstract}


\maketitle

\noindent 
Any realistic quantum system inevitably interacts with its
environment, therefore, the theory of open quantum systems and their
dynamical features is of particular importance \cite{Breuer,Weiss}.
Actually, this problem  attracts nowadays increasing attention  due
to the growing interest in controlling quantum systems and
applications in modern quantum technologies such as quantum
communication, cryptography and computation \cite{QIT}.

For several decades the popular Markovian approximation which does
not take into account memory effects was successfully studied and
applied in a variety of problems \cite{Breuer,Weiss}. However,
recent investigations in quantum information and recent
technological progress calls for truly non-Markovian approach. In
the last few years many analytical methods and numerical techniques
have been developed to treat non-Markovian processes in quantum
optics, solid state physics and quantum information
\cite{Wilkie,Budini,B-2004,Wodkiewicz,Lidar,Maniscalco1,Maniscalco2,Maniscalco-09,KR,KR-last,B,Francesco,DAS,PRL}.
Moreover, several measures of non-Markovianity were proposed and
intensively studied
\cite{Wolf,Breuer-09,Breuer-10a,Breuer-10b,Plenio}.

The most general form of local in time Master Equation reads as
follows
 \begin{equation}\label{G}
    \frac{d\rho(t)}{dt} = \mathcal{L}(t,t_0)\, \rho(t)  \ , \ \ \ \  \rho(t_0)=\rho_0\
    ,
\end{equation}
where ${\cal L}(t,t_0)$ is a local generator which  depends not only
upon the current time `$t$' but in principle it might depend upon
the initial point `$t_0$'. It is clear that dependence on `$t_0$'
introduces an effective memory. The system does remember when the
evolution begun. We call the evolution governed by (\ref{G})
Markovian if and only if ${\cal L}(t,t_0)$ does not depend on
`$t_0$'. Otherwise the evolution is non-Markovian. It is clear that
$\mathcal{L}(t,t_0)$ is defined for $t\geq t_0$ only. Note however,
that in the Markovian case $\mathcal{L}_{\rm M}(t)$ is defined in
principle for all $t\in (-\infty,\infty)$. Any solution to (\ref{G})
gives rise to the dynamical map $\Lambda(t,t_0)$ defined by $\rho(t)
= \Lambda(t,t_0)\rho_0$. Clearly $\Lambda(t,t_0)$ itself satisfies
the following equation
\begin{equation}\label{G1}
    \frac{d}{dt}\Lambda(t,t_0) = {\cal L}(t,t_0) \Lambda(t,t_0)  \ , \ \ \ \  \Lambda(t_0,t_0)=\oper\
    ,
\end{equation}
where $\oper$ denotes an identity map. A particular class of
$\mathcal{L}(t,t_0)$ is provided by the homogeneous generators, i.e.
when $\mathcal{L}$ depends on `$t-t_0$' only. Hence, the evolution
is governed by the 1-parameter family $\mathcal{L}(\tau)$  defined
for $\tau \geq 0$. It is clear that in this case the evolution is
homogeneous as well, that is
 $   \Lambda(t+T,t_0+T) = \Lambda(t,t_0)\,$
for arbitrary $T$, and hence one may define a 1-parameter family of
dynamical maps $\Lambda(t):= \Lambda(t,0)$. Actually, one usually
fixes $t_0=0$ from the very beginning and considers
\begin{equation}\label{G1-H}
    \frac{d}{dt}\Lambda(t) = {\cal L}(t) \Lambda(t)  \ , \ \ \ \  \Lambda(0)=\oper\
    .
\end{equation}
 We stress that
it could be done only in the homogeneous case and usually it is
referred as a time convolutionless (TCL) approach
\cite{TCL,BKP,Cresser}.


 A solution $\Lambda(t,t_0)$
to (\ref{G1}) is defined by the following formula
\begin{equation}\label{T-local}
    \Lambda(t,t_0) = {\rm T}\, \exp\left(\int_{t_0}^t {\cal L}(\tau,t_0)d\tau\right)\ ,
\end{equation}
where T stands for the chronological operator. In the homogeneous
case it simplifies to
\begin{equation}\label{T-local-H}
    \Lambda(t,t_0) = {\rm T}\, \exp\left(\int_{0}^{t-t_0} {\cal L}(\tau)d\tau\right)\ ,
\end{equation}
which clearly shows that $\Lambda(t,t_0)$ depends on `$t-t_0$'.
 We stress that the formula (\ref{T-local}) has only a formal
character since in general the evaluation of T-product is not
feasible. Recall, that this formula simplifies if
$\mathcal{L}(t,t_0)$ defines mutually commuting family, i.e.
$[\mathcal{L}(t,t_0),\mathcal{L}(u,t_0)]=0$ for all $t,u\geq t_0$.
In this case  T-product drops out from (\ref{T-local}).

A solution $\Lambda(t,t_0)$ defines a legitimate quantum dynamics if
and only if $\Lambda(t,t_0)$ is completely positive and trace
preserving (CPT) for all $t \geq t_0$. Now comes the natural
question: how to characterize the properties of $\mathcal{L}(t,t_0)$
which guarantee that $\Lambda(t,t_0)$ corresponds to the legitimate
quantum dynamics. These conditions are well known in the Markovian
case: a solution to
\begin{equation}\label{M}
    \frac{d}{dt}\Lambda(t,t_0) = {\cal L}_{\rm M}(t) \Lambda(t,t_0)  \ , \ \ \ \  \Lambda(t_0,t_0)=\oper\
    ,
\end{equation}
is CPT if and only if the time dependent generator has the following
Lindblad representation  \cite{Lindblad,Gorini,Alicki}
\begin{equation}\label{}
 \mathcal{L}_{\rm M}\rho = -i [H,\rho] + \sum_\alpha \gamma_\alpha \Big( V_\alpha
 \rho V_\alpha^\dagger - \frac 12 \{
 V_\alpha^\dagger V_\alpha,\rho\} \Big) \ ,
\end{equation}
where $H=H(t)$ stands for the effective time-dependent Hamiltonian
and $V_\alpha=V_\alpha(t)$ are time-dependent Lindblad (or noise)
operators. The time dependent coefficients $\gamma_\alpha$ satisfy
$\gamma_\alpha(t) \geq 0$ and encode the information about
dissipation and/or decoherence of the system.


Let us observe that a family $\Lambda(t,t_0)$ of CPT maps  may be
represented by
\begin{equation}\label{}
    \Lambda(t,t_0) = e^{Z(t,t_0)}  \ ,
\end{equation}
where $Z(t,t_0)$ has a Lindblad representation for all $t\geq t_0$.
The  price we pay for this simple representation is that $Z(t,t_0)$
might be highly singular. It is clear that formally $Z(t,t_0)$ is
defined as a logarithm of $\Lambda(t,t_0)$ and hence it is not
uniquely defined ($\log$ has an infinite number of branches).
Moreover, one always meets problems when $\Lambda(t,t_0)$ possesses
eigenvalues belonging to the cut of $\log$, cf. discussion in
\cite{Wolf}. For example a CPT map $\Lambda\rho = \sigma_z \rho
\sigma_z$ cannot be represented by $\Lambda= e^Z$. Note, however,
that $\Lambda$ may be considered as a limit of $\Lambda(t)=
e^{Z(t)}$, with $Z(t) = -\log(\cos t) L_0$, and $L_0\rho = \sigma_z
\rho \sigma_z - \rho$ is a legitimate Lindblad generator for $t\in
[0,\pi/2)$. One has $\Lambda = \lim_{t\rightarrow \pi/2} \Lambda(t)$
(see discussion in \cite{PRL}).

Note, that condition $\Lambda(t_0,t_0)=\oper$ is equivalent to
$Z(t_0,t_0)=0$ which is guarantied by
\begin{equation}\label{ZX}
    Z(t,t_0) = \int_{t_0}^t X(u,t_0)\, du\ ,
\end{equation}
and hence the solution has the following form
\begin{equation}\label{X-local}
    \Lambda(t,t_0) = \exp\left(\int_{t_0}^t X(\tau,t_0)d\tau\right)\
    .
\end{equation}
Note, that contrary to (\ref{T-local}) the above formula does not
contain chronological T-product. The corresponding generator
$\mathcal{L}(t,t_0)$ is defined by \cite{PRL}
\begin{equation}\label{}
\mathcal{L}(t,t_0) = \frac{d}{dt}\Lambda(t,t_0)\cdot
\Lambda(t,t_0)^{-1}\ ,
\end{equation}
where $\Lambda(t,t_0)^{-1}= e^{-Z(t,t_0)}$ denotes the inverse of
$\Lambda(t,t_0)$. Note, that $\Lambda(t,t_0)^{-1}$ is  not
completely positive, hence can not describe quantum evolution
backwards in time, unless   $\Lambda(t,t_0)$ is unitary or
anti-unitary. Now, to compute $d\Lambda(t,t_0)/dt$ one uses well
known formula \cite{Wilcox}
\begin{eqnarray}\label{}
    \frac{d}{dt}\, e^{A(t)} &=& \int_0^1 e^{sA(t)}\, \dot{A}(t)\, e^{(1-s)A(t)}\,
    ds \ ,
\end{eqnarray}
where $A(t)$ is an arbitrary (differentiable) family of operators,
and $\dot{A} = dA/dt$.
 Hence
\begin{equation}\label{}
\frac{d}{dt}\, e^{Z(t,t_0)} = \mathcal{L}(t,t_0) e^{Z(t,t_0)}\ ,
\end{equation}
 where
\begin{equation}\label{MAIN}
     \mathcal{L}(t,t_0) = \int_0^1 e^{sZ(t,t_0)}\, X(t,t_0)\, e^{-sZ(t,t_0)}\,     ds
\ .
\end{equation}
This is the main result of our Letter. It proves that each
legitimate generator $\mathcal{L}(t,t_0)$ of quantum evolution
governed by the Master Equation (\ref{G}) has the form defined by
(\ref{MAIN}), where $Z(t,t_0)$ has a Lindblad representation for
each $t\geq t_0$, and $X(t,t_0)$ is defined in (\ref{ZX}). Hence the
construction of a legitimate generator is pretty simple: each family
of Lindblad operators $Z(t,t_0)$, with $Z(t_0,t_0)=0$, gives rise
via (\ref{MAIN}) to the corresponding prescription for
$\mathcal{L}(t,t_0)$. Nevertheless, the formula (\ref{MAIN}) is
highly nontrivial and the computation of $\mathcal{L}(t,t_0)$ out of
$Z(t,t_0)$ might be highly complicated. This is the price we pay for
the simple representation of evolution  (\ref{X-local}). Hence, we
have a kind of complementarity: either one uses T-product formula
(\ref{T-local}) with relatively simple generator or one avoids
T-product in (\ref{X-local}) but uses highly nontrivial generator
(\ref{MAIN}). The advantage of our approach is that one knows how to
construct generator (in practice it might be complicated) giving
rise to the legitimate quantum dynamics.

Let us observe that in the special case when $X(t,t_0)$ mutually
commute, i.e. $[X(t,t_0),X(u,t_0)]=0$ for all $t,u\geq t_0$, the
formula (\ref{MAIN}) reduces to
 $   \mathcal{L}(t,t_0) = X(t,t_0)\,$.
Hence, a commuting family $\mathcal{L}(t,t_0)$
defines a legitimate generator if and only if
 $   Z(t,t_0) = \int_{t_0}^t \mathcal{L}(u,t_0)\, du\,$
has a Lindblad representation for all $t\geq t_0$. In the
noncommutative case this simple criterion is no longer true.

The characteristic feature of the Markovian evolution governed by
(\ref{M}) is that $\Lambda(t,t_0)$ satisfies  local composition law
\begin{equation}\label{COMP-t}
   \Lambda(t,s) \cdot \Lambda(s,t_0) = \Lambda(t,t_0)\ ,
\end{equation}
for $t\geq s\geq t_0$. Actually, this property is guarantied by the
intricate action of T-product in the formula (\ref{T-local}). Now,
changing the representation from (\ref{T-local}) into
(\ref{X-local}) the validity of composition law is no longer
visible. The formula  (\ref{COMP-t}) implies
\begin{equation}\label{}
    e^{Z(t,u)} \cdot e^{Z(u,t_0)} = e^{Z(t,t_0)} \ ,
\end{equation}
for $t\geq s\geq t_0$ Clearly, in the commutative one has simply
\begin{equation}\label{}
    Z(t,u) + Z(u,t_0) = Z(t,t_0)\ .
\end{equation}
Note however that when $Z(t,t_0)$ do not commute, the
Baker-Campbell-Hausdorff formula $e^A e^B = e^C$, with
\begin{equation*}\label{}
    C = A+B + \frac 12 [A,B] + \frac{1}{12} \Big( [A,[A,B]] -
    [B,[A,B]] \Big) + \ldots \ ,
\end{equation*}
provides highly nontrivial condition upon $Z(t,t_0)$. It shows that
knowing legitimate $Z(t,t_0)$  one does not know immediately wether
the corresponding dynamics is Markovian or not. Only applying
(\ref{MAIN}) one can check wether $\mathcal{L}(t,t_0)$ does, or does
not, depend on `$t_0$' and hence infer about Markovianity. This
problem simplifies in the homogeneous case: now the evolution is
never Markovian unless $Z(t,t_0) =(t-t_0) L_0$, i.e.
$\mathcal{L}(t,t_0) = L_0$, where $L_0$ denotes the time independent
Markovian generator.

It is clear that for a general family of Lindblad generators
$Z(t,t_0)$ computation of $\mathcal{L}(t,t_0)$ via (\ref{MAIN}) is
rather untractable. However, us usual, there is a class of
$Z(t,t_0)$ for which the general problem simplifies considerably.
Consider a special class of $Z(t,t_0)$ defined via (\ref{ZX}) by the
following family
\begin{equation}\label{XLL}
    X(t,t_0) = a_1(t,t_0) \mathcal{L}_1 + \ldots + a_N(t,t_0)
    \mathcal{L}_N \ ,
\end{equation}
where $\mathcal{L}_1,\ldots,\mathcal{L}_N$ are time independent
Lindblad generators. One has
\begin{equation}\label{ZLL}
    Z(t,t_0) = A_1(t,t_0) \mathcal{L}_1 + \ldots + A_N(t,t_0)
    \mathcal{L}_N \ ,
\end{equation}
where  $A_k(t,t_0) = \int_{t_0}^t  a_k(u,t_0)du$. Now, $Z(t,t_0)$
has a Lindblad representation iff $A_k(t,t_0) \geq 0$. Let us
observe that if $\mathcal{L}_1,\ldots,\mathcal{L}_N$ close a Lie
algebra, i.e. $[\mathcal{L}_j,\mathcal{L}_j]= \sum_{k=1}^N c^k_{ij}
\mathcal{L}_k$, then using well known Lie algebraic methods one can
easily compute $\mathcal{L}(t,t_0)$ out of (\ref{MAIN}) and gets
\begin{equation}\label{XLL}
    \mathcal{L}(t,t_0) = b_1(t,t_0) \mathcal{L}_1 + \ldots + b_N(t,t_0)
    \mathcal{L}_N \ ,
\end{equation}
where the functions $b_k(t,t_0)$ are uniquely defined by
$a_k(t,t_0)$ and the structure constants $c^k_{ij}$. Actually, any
set $\{\mathcal{L}_1,\ldots,\mathcal{L}_N\}$ of Lindblad generators
may be always completed to close a Lie algebra. It follows from the
fact that a set of Lindblad generators belong to the Lie algebra
corresponding to the Lie group of linear maps preserving
hermiticity. Note, that if $[\mathcal{L}_i,\mathcal{L}_j]=0$, i.e.
the corresponding Lie algebra is commutative, then
$b_k(t,t_0)=a_k(t,t_0)$, that is, $\mathcal{L}(t,t_0)= X(t,t_0)$.

\begin{example}[Commutative case]   \label{E-PURE}
{\em  Consider the following pure decoherence model defined by the
following time dependent Hamiltonian  $H(t)=H_R(t) + H_S(t) +
H_{SR}(t)$, where $H_R(t)$ is the reservoir Hamiltonian, $H_S(t) =
\sum_n \epsilon_n(t) P_n \; (P_n=|n\>\< n|)$ the system Hamiltonian
and
\begin{equation}\label{}
    H_{SR}(t) = \sum_n P_n \ot B_n(t)
\end{equation}
the interaction part, $B_n=B_n^\dagger$ being reservoirs operators.
The initial product state $\rho \ot \omega$ evolves according to the
unitary evolution $U(t,t_0) (\rho \ot \omega) U(t,t_0)^\dagger$ and
by partial tracing with respect to the reservoir degrees of freedom
one finds for the evolved system density matrix
\begin{eqnarray}\label{rho0}
    \rho(t) = \Lambda(t,t_0)\rho 
    = \sum_{n,m} c_{mn}(t,t_0)P_m\rho P_n \ ,
\end{eqnarray}
where $c_{mn}(t,t_0) = {\rm Tr}[ U_m(t,t_0)\,\omega\,
U_n(t,t_0)^\dagger]$, with $U_n(t,t_0) = {\rm T} \exp[ -
i\int_{t_0}^t Y_n(\tau)d\tau]$, and $Y_n(\tau) = \epsilon_n(\tau)
\mathbb{I}_R + H_R(\tau) + B_n(\tau)\,$ being time dependent
reservoir operators. Note that the matrix $c_{mn}(t,t_0)$ is
semi-positive definite and hence (\ref{rho0}) defines the Kraus
representation of the completely positive map $\Lambda(t,t_0)$. Note
that $\Lambda(t,t_0)$ defines a commutative family of maps and hence
one easily finds for the corresponding generator
\begin{equation}\label{L-pure}
\mathcal{L}(t,t_0)\,\rho = \sum_{n,m} \alpha_{mn}(t,t_0) P_m \rho
P_n\ ,
\end{equation}
where the functions $\alpha_{mn}(t,t_0)$ are defined by
$\alpha_{mn}(t,t_0)= \dot{c}_{mn}(t,t_0)/c_{mn}(t,t_0)$. Note that
if ${\rm dim}\mathcal{H}_S=2$, then $c_{11}=c_{22}=1\,$, and $c_{12}
= \gamma$ with $|\gamma|\leq 1$. One easily finds for the local
generator
\begin{equation}\label{L-com}
    \mathcal{L}(t,t_0)\rho = ib_1(t,t_0) [\sigma_z,\rho] -
    b_2(t,t_0) [\sigma_z \rho \sigma_z - \rho]\ ,
\end{equation}
with $b_1 = {\rm Im} (\dot{\gamma}/2\gamma)$ and  $b_2 = {\rm Re}
(\dot{\gamma}/2\gamma)$.  Note that this dynamics is homogeneous if
and only if the Hamiltonian of $S+R$ is time independent.}
\end{example}

\begin{example}[Noncommutative case] \label{EX-2}
 {\em Let us consider a  simple example of exactly solvable dynamics of 2-level
system defined by the  following  homogenous family of operators
$X(t):= X(t,0)$
\begin{equation}\label{}
    X(t) = a_1(t) \mathcal{L}_1 + a_2(t) \mathcal{L}_2 \ ,\ \ \ t
    \geq 0 \ ,
\end{equation}
where the Markovian generators $\mathcal{L}_1,\mathcal{L}_2$ are
defined by
\begin{eqnarray*}
  \mathcal{L}_1\rho &=& \sigma^+ \rho \sigma^- - \frac 12 \{ \sigma^-\sigma^+,\rho\} \ , \\
  \mathcal{L}_2\rho &=& \sigma^- \rho \sigma^+ - \frac 12 \{ \sigma^+\sigma^-,\rho\} \ , \\
\end{eqnarray*}
and $\sigma^+ = |1\>\<2|\, $, $\, \sigma^- = |2\>\<1|\,$
are the standard raising and lowering qubit operators
($\{|1\>,|2\>\}$ denotes  an orthonormal basis in the qubit Hilbert
space). Since $\mathcal{L}_1$ and $\mathcal{L}_2$ do not commute
$X(t)$ defines a noncommutative family. Clearly, one may add to
$X(t)$ a commutative part (\ref{L-com}) which does commute with
$X(t)$ 
and hence do not change qualitative features of dynamic. For
simplicity we consider only simplified version which is essential
for non-commutativity.

 The time dependent parameters $a_1(t)$ and $a_2(t)$ are
arbitrary but real.   Following our construction one has
\begin{equation}\label{}
    Z(t) =  A_1(t) \mathcal{L}_1 + A_2(t) \mathcal{L}_2 \ ,
\end{equation}
where $A_k(t)= \int_0^t a_k(u) du$.
 Hence, the formula
 $   \Lambda(t) = e^{Z(t)}\,$
defines CPT map for all $t \geq 0$ if and only if
\begin{equation}\label{AB>0}
    A_1(t) \geq 0 \ , \ \ \ A_2(t) \geq 0\ .
\end{equation}
Now, let us apply our basic formula (\ref{MAIN}) to find the
corresponding generator $\mathcal{L}(t)$. Observing that
$\mathcal{L}_1$ and $\mathcal{L}_2$ close a Lie algebra
$    [\mathcal{L}_1,\mathcal{L}_2] = \mathcal{L}_1 -
\mathcal{L}_2\,$,
one easily finds
\begin{equation}\label{}
    \mathcal{L}(t)  = b_1(t) \mathcal{L}_1 + b_2(t) \mathcal{L}_2 \ ,
\end{equation}
where
\begin{eqnarray}
  b_1(t) = a_1(t) + f(t) \ ,  \ \ \ \
  b_2(t) = a_2(t) - f(t)\ ,
\end{eqnarray}
and the time dependent function $f(t)$ reads as follows
\begin{equation}\label{f}
    f(t) = \frac{W(t)}{A(t)} \left( 1 + \frac{e^{-A(t)}
    -1}{A(t)} \right) ,
\end{equation}
where the Wronskian $W(t) = A_1(t)a_2(t) - A_2(t)a_1(t)$, and $A(t)
= A_1(t) + A_2(t)$. Note, that $f(t)=0$ (for all $t \geq 0$) if and
only if $W(t)=0$, i.e. functions $a_1(t)$ and $a_2(t)$ are linearly
dependent. If this is the case one has $a_2(t) = \lambda a_1(t)$,
and $X(t) = a_1(t) ( \mathcal{L}_1 + \lambda \mathcal{L}_2)$ defines
a commutative family. In this case one has $b_k(t) = a_k(t)$ and
hence $\mathcal{L}(t)=X(t)$. In the general noncommutative case one
has for  the integral
\begin{equation}\label{}
\int_0^{t} \mathcal{L}(\tau) d\tau = B_1(t) \mathcal{L}_1 + B_2(t)
\mathcal{L}_2 \ ,
\end{equation}
with   $ B_1(t) = A_1(t) + F(t)\,$,  $B_2(t) = A_2(t) - F(t)\,$, and
$F(t) = \int_0^t f(u)du$. Note, that $B_1(t) + B_2(t) = A(t) \geq
0$. However, contrary to the commutative case, there is no need that
both $B_1(t)$ and $B_2(t)$ are positive. It shows that integral
$\int_0^{t} \mathcal{L}(\tau) d\tau$ needs not have a Lindblad
representation. Note, that
\begin{equation}\label{}
    \mathcal{L}(t) = X(t) + f(t)[\mathcal{L}_1-\mathcal{L}_2]\ ,
\end{equation}
which clearly shows that the last term
`$f(t)[\mathcal{L}_1-\mathcal{L}_2]$' destroys the Lindblad
structure of $\int_0^t \mathcal{L}(\tau)d\tau$. It proves the
intricate action of T-product:
\begin{equation}\label{}
    {\rm T} \exp{\left( \int_0^t \Big\{ X(\tau) +
    f(\tau)[\mathcal{L}_1-\mathcal{L}_2]\Big\} d\tau
    \right)}
    = e^{Z(t)}\ ,
\end{equation}
that is, chronological product simply washes out the unwanted term
`$f(t)[\mathcal{L}_1-\mathcal{L}_2]$'. Eventually, one easily shows
(using standard algebraic methods, e.g. \cite{Wei}) that
\begin{equation}\label{Z-nu}
    e^{Z(t)} = e^{\ln\nu_1(t) \mathcal{L}_1} \cdot  e^{\ln\nu_2(t)
    \mathcal{L}_2}\ ,
\end{equation}
where (skipping time dependence)
\begin{equation}\label{}
    \nu_1 = \frac{A}{A_1 e^{-A} + A_2} \ , \ \ \ \nu_2 = \frac{A_1 +
    A_2 e^A}{A} \ .
\end{equation}
Note that $\nu_1 \nu_2 = e^{A}$. One has $\nu_k(t) \geq 1$, and
hence (\ref{Z-nu}) gives another representation of dynamical map as
a composition of two completely positive maps generated by
$\mathcal{L}_1$ and $\mathcal{L}_2$.

 }
\end{example}

In conclusion, we proposed a complete treatment of a local in time
dynamics of open quantum systems based on the Master Equation
(\ref{G}). We provided a general representation of the local
generator -- formula (\ref{MAIN}) -- which generalizes well known
Lindblad representation for the Markovian dynamics. We stress that
any local generator $\mathcal{L}(t,t_0)$ may be constructed via
(\ref{MAIN}) by a suitable choice of the Lindblad family $Z(t,t_0)$.
However, the problem of necessary and sufficient condition for
$\mathcal{L}(t,t_0)$ which guarantee that $\Lambda(t,t_0)$ is CPT is
still open. Only, if $\mathcal{L}(t,t_0)$ defines a commutative
family, these conditions reduce to a simple requirement  that
$\int_{t_0}^t \mathcal{L}(u,t_0)du$ has a Lindblad form for $t\geq
t_0$.

\noindent {\em Acknowledgments}. This work was partially supported
by the Polish Ministry of Science and Higher Education Grant No
3004/B/H03/2007/33.

\end{document}